\documentstyle[preprint,aps]{revtex}
\tightenlines
\begin{document}
\draft
\title{$^{13}_{\Lambda}$C hypernucleus studied with the 
$^{13}$C($K^{-}$, $\pi^{-}\gamma$) reaction}
\author{H. Kohri, 
        S. Ajimura, 
        H. Hayakawa, 
        T. Kishimoto, 
        K. Matsuoka, 
        S. Minami, 
        Y. S. Miyake, \\
        T. Mori, 
        K. Morikubo, 
        E. Saji, 
        A. Sakaguchi, 
        Y. Shimizu, and 
        M. Sumihama}
\address{Department of Physics, Osaka University, 
         Toyonaka, Osaka, 560-0043, Japan}
\author{R. E. Chrien, 
        M. May, 
        P. Pile, 
        A. Rusek, and 
        R. Sutter}
\address{Brookhaven National Laboratory, 
         Upton, New York 11973, USA}

\author{P. M. Eugenio, 
        G. Franklin, 
        P. Khaustov, 
        K. Paschke, 
        B. P. Quinn, and 
        R. A. Schumacher}
\address{Department of Physics, Carnegie Mellon University, 
        Pittsburgh, PA 15213, USA}

\author{J. Franz}
\address{Department of Physics, University of Freiburg, 
         D79104 Freiburg, Germany}

\author{T. Fukuda, 
        H. Noumi, and 
        H. Outa}
\address{High Energy Accelerator Research Organization (KEK), 
        Tsukuba, Ibaragi, 305-0801, Japan}

\author{L. Gan, 
        L. Tang, and 
        L. Yuan}
\address{Department of Physics, Hampton University, Hampton, 
         VA 23668, USA}

\author{J. Nakano, 
        T. Tamagawa, and 
        K. Tanida}
\address{Department of Physics, University of Tokyo, 
         Tokyo 113-0033, Japan}

\author{R. Sawafta}
\address{Department of Physics, North Carolina A T State 
         University, Greensboro, NC 27411, USA}

\author{H. Tamura}
\address{Department of Physics, Tohoku University, Sendai 
         980-8578, Japan}

\author{H. Akikawa}
\address{Department of Physics, Kyoto University, 
         Kyoto 606-8502, Japan}
\date{\today}
\maketitle
\begin{abstract}
The $^{13}_{\Lambda}$C hypernucleus was studied by measuring $\gamma$ rays 
in coincidence with 
the $^{13}$C($K^{-}$, $\pi^{-}$) reaction. 
$\gamma$ rays from the 1/2$^{-}$ and 3/2$^{-}$ states, which 
are the partners of the spin-orbit doublet states with 
a predominant configuration 
of [$^{12}$C$_{g.s.}$(0$^{+}$) $\otimes$ $p_{\Lambda}$], 
to the ground state were measured. 
The splitting of the states was found to be 
$\Delta$E(1/2$^{-}-$3/2$^{-}$) = 
+152$\pm$54(stat)$\pm$36(syst) keV. 
This value is 20$\sim$30 times smaller than that of single particle 
states in nuclei around this mass region. 
The $j_{\Lambda}$=$\ell_{\Lambda}-$1/2 (($p_{1/2}$)$_{\Lambda}$) state 
appeared higher in energy, as in normal nuclei. 
The value gives new insight into the $YN$ interaction. 
The excitation energies of the 1/2$^{-}$ and 3/2$^{-}$ states were 
obtained as 
10.982$\pm$0.031(stat)$\pm$0.056(syst) MeV and 
10.830$\pm$0.031(stat)$\pm$0.056(syst) MeV, respectively. 
We also observed $\gamma$ rays from the 3/2$^{+}$ state, which has a 
[$^{12}$C(2$^{+}$) $\otimes$ $s_{\Lambda}$] configuration, 
to the ground state in $^{13}_{\Lambda}$C. 
The excitation energy of the 3/2$^{+}$ state was obtained as 
4.880$\pm$0.010(stat)$\pm$0.017(syst) MeV. 
Nuclear $\gamma$ rays with energies of 4.438 and 15.100 MeV had similar 
yields, which suggests that a quasi-free knockout of a $\Lambda$ 
particle is dominant in highly excited regions. 
\end{abstract}

\narrowtext

\section{INTRODUCTION}
\label{sec:level1}

The introduction of the spin-orbit ($\ell s$) force, which is 
a short range interaction, resulted in 
great successes of the nuclear shell model. 
Before this introduction, the central force described by the harmonic
oscillator had been unable to explain magic numbers except for 2, 
8, and 20. 
The $\ell s$-force clearly explained not only the magic numbers 
but also many prominent nuclear properties. 
The $\ell s$-splitting of single nucleon states is as large as that of 
the major shell spacing and plays an essential role in nuclear 
physics. 

As for $\Lambda$-nucleus interactions in $\Lambda$-hypernuclei, 
the observation of $\Lambda$ single particle states, first by the 
($K^{-}$, $\pi^{-}$) reaction \cite{Chrien1}
and then later by the ($\pi^{+}$, $K^{+}$) 
reaction \cite{Pile1,Hasegawa1}, 
clarified the gross structure of the interactions. 
The central force was found to be roughly 2/3 of that of a nucleon. 
In the naive quark model a $\Lambda$ particle is composed of u, d, and 
s quarks and the s (strange) quark is considered to contribute little 
to the nuclear force. 
In the one boson exchange (OBE) model it is understood that the absence 
of the long-range one pion exchange contribution makes the central 
force smaller. 
On the other hand, spin-dependent forces are not well known. 
Especially, the $\ell s$-force in a $\Lambda$-nucleus is 
considered to be much smaller than that in a nucleon-nucleus, 
although no experiment has given a conclusive value. 
Measurement of the $\Lambda$-nucleus $\ell s$-splitting has been one 
of major objects in the study of hypernuclei for more than two 
decades. 

This paper is a follow-up to our recently-published 
paper \cite{Ajimura0} which was confined to the derivation of the 
$\ell s$-splitting. 
In this paper, details of the experiment and analysis 
are provided and spectroscopic results for the $^{13}_{\Lambda}$C 
hypernucleus are discussed. 

\subsection{Experimental studies}
\label{sec:level2}

The first indication of a small $\Lambda$-nucleus 
$\ell s$-splitting was given by the study of the 
$^{16}$O($K^{-}$, $\pi^{-}$)$^{16}_{\Lambda}$O reaction at the CERN 
Proton Synchrotron (PS)~\cite{Bruckner} in 1978. 
They observed peaks with major configurations of 
[($p_{1/2}$)$^{-1}_{n}$($p_{1/2}$)$_{\Lambda}$]$_{0^{+}}$ 
and [($p_{3/2}$)$^{-1}_{n}$($p_{3/2}$)$_{\Lambda}$]$_{0^{+}}$. 
The splitting of the two peaks was about 6 MeV which is quite 
close to 6.18 MeV corresponding to a splitting between the 
($p_{1/2}$)$^{-1}_{n}$ and ($p_{3/2}$)$^{-1}_{n}$ states in $^{15}$O 
(core nucleus of $^{16}_{\Lambda}$O). 
This result suggested that the $\Lambda$-$^{15}$O $\ell s$-force was 
extremely small. 
Br$\ddot{\rm u}$ckner $et$ $al$. gave an upper limit of 0.3 MeV for 
the $\ell s$-splitting. 
However a detailed theoretical analysis gave 0.8$\pm$0.7 MeV for 
the splitting \cite{Bouyssy1}. 
Consequently the constraint on the splitting was weakened. 
Later May $et$ $al$. measured an energy shift of a peak 
at E$_{x}\sim$10.4 MeV between 0$^{\circ}$ and 15$^{\circ}$ by the 
$^{13}$C($K^{-}$, $\pi^{-}$)$^{13}_{\Lambda}$C reaction at the 
AGS of BNL \cite{May1}. 
Predominant configurations of the peak were 
[($p_{1/2}$)$^{-1}_{n}$($p_{1/2}$)$_{\Lambda}$]$_{1/2^{-}}$ 
at 0$^{\circ}$ and 
[($p_{1/2}$)$^{-1}_{n}$($p_{3/2}$)$_{\Lambda}$]$_{3/2^{-}}$ 
at 15$^{\circ}$. 
They obtained 0.36$\pm$0.3 MeV for the splitting of the 
1/2$^{-}$ and 3/2$^{-}$ states in $^{13}_{\Lambda}$C. 
Another study was the observation of $\gamma$ rays from 
$^{9}_{\Lambda}$Be excited by the $^{9}$Be($K^{-}$, $\pi^{-}$) 
reaction by using NaI detectors \cite{May2}. 
A peak was observed at 3.079$\pm$0.040 MeV in the $\gamma$ ray 
spectrum and the peak was considered to be due to $\gamma$ rays from 
unresolved 3/2$^{+}$ and 5/2$^{+}$ states, which have a configuration 
of [$^{8}$Be(2$^{+}$) $\otimes$ ($s_{1/2}$)$_{\Lambda}$], 
to the ground state (GS) in $^{9}_{\Lambda}$Be. 
The width of the peak suggested that the splitting was less than 
0.1 MeV, although the possibility that $\gamma$ rays from either 
state were missing was not completely excluded. 

On the other hand, data suggesting larger $\ell s$-splittings 
appeared recently. 
Nagae $et$ $al$. observed a series of peaks 
considered to be associated with states with a $\Lambda$ particle 
in the $s$, $p$, $d$, and $f$-orbits by the 
$^{89}$Y($\pi^{+}$, $K^{+}$)$^{89}_{\Lambda}$Y reaction 
at KEK-PS \cite{Nagae1}. 
The widths of the peaks seemed to be broader for larger $\ell$ states. 
Since the $\ell s$-splitting increases almost linearly with the 
orbital angular momentum, the peaks could be interpreted 
as unresolved $\ell s$-doublets with $V_{\ell s}\sim$6 MeV by 
using the Woods-Saxon prescription. 
This result was further supported by the re-analysis of the emulsion 
data of $^{16}_{\Lambda}$O collected by the European $K^{-}$ 
collaboration \cite{Dalitz1}. 
Dalitz $et$ $al$. assigned configurations of 
[($p_{1/2}$)$^{-1}_{n}$($p_{1/2}$)$_{\Lambda}$]$_{0^{+}}$ and 
[($p_{1/2}$)$^{-1}_{n}$($p_{3/2}$)$_{\Lambda}$]$_{2^{+}}$ 
to two peaks which were separated by 1.56$\pm$0.12 MeV. 
They derived splitting consistent with 
the $^{89}_{\Lambda}$Y data. 

\subsection{Theoretical studies}
\label{sec:level3}

Interactions between baryons have been studied by the 
meson exchange theory combined with phenomenology. 
The OBE model of the nucleon-nucleon ($NN$) interaction 
was extended to the hyperon-nucleon ($YN$) interaction with the help of 
flavor SU(3) symmetry \cite{Nagels1}. 
Effective one-body hyperon-nucleus interactions were constructed 
based on the OBE models \cite{Dover1,Yamamoto1}. 
They well reproduce the central part of the $\Lambda$-nucleus 
interaction and predict a weak $\Lambda$-nucleus $\ell s$-force. 
For example, the $\ell s$-splitting of $\Lambda$ single particle states 
in $^{13}_{\Lambda}$C was calculated to be 0.56 MeV, 
E$_{x}$=10.56 MeV for ($p_{1/2}$)$_\Lambda$ and E$_{x}$=10.00 MeV 
for ($p_{3/2}$)$_\Lambda$ \cite{Itonaga1}. 

There has been another attempt to study the short-range part 
of the $NN$ interaction from a standpoint that baryons are made of 
quarks and the short-range part of baryon-baryon interactions should 
be understood by quark and gluon exchanges. 
Morimatsu $et$ $al$. studied the $\ell s$-force between 
baryons within the framework of a nonrelativistic quark-cluster 
model \cite{Morimatsu1}. 
Since this model had a strong anti-symmetric spin-orbit ($ALS$) 
force, which was opposed to the symmetric $\ell s$-force, it 
gave a very small $\Lambda$-nucleus $\ell s$-splitting. 
Pirner and Povh also predicted zero splitting for the 
$\Lambda$-nucleus $\ell s$-splitting \cite{Pirner1}. 

Results given by the recent theoretical studies were essentially 
similar to those described above \cite{Keil1,Takeuchi1,Takeuchi2}. 
The quark models tend to predict small 
$\Lambda$-nucleus $\ell s$-splittings compared with those of the OBE 
models. 
A new experiment with a precision of better than 0.1 MeV for the $\ell 
s$-splitting was needed to distinguish the models. 

\subsection{The $^{13}_{\Lambda}$C hypernucleus}
\label{sec:level4}

The 1/2$^{-}$ and 3/2$^{-}$ states in $^{13}_{\Lambda}$C are ideal 
states to extract the $\ell s$-splitting. 
It is well known that the 1/2$^{-}$ and 3/2$^{-}$ states have 
predominant configurations of 
[$^{12}$C$_{g.s.}$(0$^{+}$) $\otimes$ ($p_{1/2}$)$_{\Lambda}$] 
and [$^{12}$C$_{g.s.}$(0$^{+}$) $\otimes$ ($p_{3/2}$)$_{\Lambda}$], 
respectively. 
The two states are the partners of the $\ell s$-doublet states. 
By virtue of the 0$^{+}$ spin of the $^{12}$C core, the energy 
difference between the states gives the $\ell s$-splitting of 
a $\Lambda$ particle in the $p$-orbit. 
Recently $\gamma$ rays from the 1/2$^{-}$ state to the GS were 
observed at 10.95 MeV by using NaI detectors \cite{May3}. 
This was the first observation of the $p_{\Lambda}$ $\rightarrow$ 
$s_{\Lambda}$ $\gamma$ ray transition. 

\section{EXPERIMENT}
\label{sec:level5}

\subsection{Principles}
\label{sec:level6}

The $\ell s$-splitting has been measured mostly by using magnetic 
spectrometers. 
The best energy resolution that magnetic spectrometers have achieved in 
the study of hypernuclei is around 2 MeV (FWHM). 
Since the $\ell s$-splitting is predicted to be 0$\sim$1 MeV, 
measurement with a precision of better than 0.1 MeV is necessary. 
In order to improve the energy resolution, we measured $\gamma$ rays 
from the 1/2$^{-}$ and 3/2$^{-}$ states at E$_{x}\sim$11 MeV to the 
GS in $^{13}_{\Lambda}$C by using NaI detectors. 
The energy resolution of the NaI detectors was about 0.35 MeV (FWHM) 
for the detection of 11 MeV $\gamma$ rays. 
It was good enough for the measurement of the 
$\ell s$-splitting with a precision of better than 0.1 MeV given 
enough statistics. 
If the splitting is larger than or close to the energy resolution of 
the NaI detectors, one should observe two peaks, or a peak 
clearly broader than the energy resolution, from which one can derive 
the splitting easily. 

Even if the splitting is much less than the energy resolution, the 
splitting can be derived as follows. 
The ($K^{-}$, $\pi^{-}$) reaction at forward angles is a unique 
way to selectively excite the 1/2$^{-}$ state in $^{13}_{\Lambda}$C 
via the $\Delta \ell$=0 transition. 
The so-called substitutional transition 
(($p_{1/2}$)$_{n}$ $\rightarrow$ ($p_{1/2}$)$_{\Lambda}$) is dominant 
at momentum transfers much smaller than the Fermi momentum. 
On the other hand the 3/2$^{-}$ state is mainly excited at larger 
angles (10$\sim$20 degrees) of the ($K^{-}$, $\pi^{-}$) reaction
via the $\Delta \ell$=2 transition. 
Angular distributions of the 
$^{13}$C($K^{-}$, $\pi^{-}$)$^{13}_{\Lambda}$C reaction for 
the 1/2$^{-}$ and 3/2$^{-}$ states at 0.9 GeV/c, calculated by 
Motoba $et$ $al$. with the distorted wave impulse approximation 
(DWIA) \cite{Motoba1}, are shown in Fig.\ref{fig:cross}. 
One can simultaneously excite both states in an experiment and 
control the ratio of the 1/2$^{-}$ and 
3/2$^{-}$ states in an analysis by utilizing the angular 
distributions. 
This method tells us not only the splitting but also which state has 
higher or lower energy. 

Based on these considerations we designed an experiment to 
measure $\gamma$ rays from the 1/2$^{-}$ and 
3/2$^{-}$ states which were 
excited by the $^{13}$C($K^{-}$, $\pi^{-}$)$^{13}_{\Lambda}$C 
reaction \cite{Kishimoto1}. 

\subsection{Experimental setup}
\label{sec:level7}

The experiment (AGS-E929) was carried out by using the D6 beam 
line \cite{Pile2} of the alternating-gradient synchrotron (AGS) 
at Brookhaven National Laboratory (BNL). 
The $K^{-}$ beam momentum was 0.93 GeV/c, or close to 0.9 GeV/c 
at the target center after energy loss, in order to maximize the 
production rate of the 1/2$^{-}$ and 3/2$^{-}$ states 
in $^{13}_{\Lambda}$C. 
The typical $K^{-}$ beam intensity was about 8 $\times$ 10$^{4}$/spill 
for 5 $\times$ 10$^{12}$ protons /spill at the primary target. 
A spill consisted of 1.4 seconds of continuous beam every 4 seconds. 
The typical ratio of ($\pi^{-}$$\mu^{-}$e$^{-}$)/$K^{-}$ was 0.3. 
The intensity and purity of the $K^{-}$ beam provided by the D6 beam line 
greatly exceeded those available elsewhere. 

The momentum of incoming beam particles was measured using information 
from a scintillator hodoscope located at the exit of the first mass 
slit in the beam line and from two drift chambers \cite{Pile2}. 
Incoming $K^{-}$ particles were identified electrically using a 
plastic scintillator (BS), an aerogel 
${\rm \check{C}}$erenkov counter, and a quartz 
${\rm \check{C}}$erenkov counter (BQC). 
The aerogel ${\rm \check{C}}$erenkov counter, which had an aerogel 
block with the refractive index of 
1.03, was installed downstream of the two drift chambers. 
The BQC, which was a total reflection type ${\rm \check{C}}$erenkov 
counter, was installed in front of the target. 
The detection efficiency for the incoming $K^{-}$ particles was 
greater than 99 \%. 
The time-of-flight between a scintillator hodoscope in the beam 
line and a timing counter at the exit of the beam line was also 
used to identify the incoming $K^{-}$ beam particles in the 
off-line analysis. 
The $K^{-}$ beam was focused on the target in the vertical direction 
by the last beam line quadrupole magnet. 
The size of the $K^{-}$ beam was 5.0 cm (FWHM) in the horizontal 
direction and 1.1 cm (FWHM) in the vertical direction at the target. 

The detector configuration around the target is shown in 
Fig.\ref{fig:aroundtgt}. 
The target had four cells with inner dimensions of 6 cm wide, 
1.5 cm high, and 3 cm thick and containing $^{13}$C benzene enriched 
to 99 \%. 
Each target cell was made of quartz with a wall thickness of 1 mm. 
Accordingly, the target had some contamination of O and Si nuclei. 
In the $^{13}$C benzene, laser dyes were added to make the 
target scintillate. 
It is known that the ($K^{-}$, $\pi^{-}$) reaction is strongly
contaminated by $K^{-}$ in-flight decays. 
The production of $^{13}_{\Lambda}$C hypernuclei was identified by 
the light output of the scintillating target, and it was possible 
to discriminate against $K^{-}$ decay events. 
Without using this method, the 
production of $^{13}_{\Lambda}$C hypernuclei could not be clearly 
observed at certain angles where the kinematics of the $K^{-}$ 
decays was the same as that of the 
$^{13}$C($K^{-}$, $\pi^{-}$)$^{13}_{\Lambda}$C reaction. 
The tagging efficiency for the $^{13}_{\Lambda}$C weak decay was 
greater than 80 \% by selecting events with higher light output. 
Four plastic scintillators (DEC) were installed above and below the 
target to increase the tagging efficiency for the weak decay by 
detecting decay particles from near the surface of the 
scintillating target. 
The size of each scintillator was 15 cm $\times$ 24 cm $\times$ 1.5 cm. 
Event selection using the DEC and the scintillating target is 
described in Sec.\ref{sec:level10} and details are given 
in Ref.\cite{Ajimura1,Kohri1}. 

72 NaI detectors (6$\times$6 above and 6$\times$6 below the 
target) were installed at a distance of 10.5 cm from the target center. 
The segmentation was important to withstand high counting rates and 
was also quite convenient to correct for $\gamma$ ray Doppler shift. 
The size of each NaI crystal was 6.5 cm $\times$ 6.5 cm $\times$ 30 cm. 
A 2$''$ PM tube (Hamamatsu H1161), which was covered with triple 
magnetic shield cases, was connected to the crystal through a silicon 
rubber disk and a light guide. 
The typical counting rate was 5 $\times$ 10$^{3}$ counts/spill. 
An 11 MeV $\gamma$ ray is energetic enough to make an electromagnetic 
shower in a NaI crystal. 
In most cases one detector, called the central detector,
received the main energy deposit and neighboring detectors had smaller 
energy deposits because only 511 keV annihilation $\gamma$ rays 
dominantly escaped from the central detector. 
For increasing the full energy peak efficiency, the energy deposit of 
0.1$\sim$1.2 MeV in each of three detectors was added up. 
Since the energy resolution was mainly determined by that of the 
central detector, NaI detectors with good energy resolution were 
installed in the center of the NaI array. 
Eight plastic scintillators (Charge Veto) were installed on 
the surface of the NaI array to reject the charged particle 
background. 
The size of each scintillator was 42 cm $\times$ 10 cm $\times$ 1 cm. 
When both a charged particle and a $\gamma$ ray passed through 
the same element of the Charge Veto, it rejected not only the charged 
particle but also the $\gamma$ ray. 
Thus, the detection efficiency of the NaI detectors for $\gamma$ rays 
was decreased by 10 \%. 
The total detection efficiency of the NaI detectors for 11 MeV 
$\gamma$ rays was 4.5 \% most of which was determined by the geometry. 

$\pi^{-}$ particles were identified by using an aerogel 
${\rm \check{C}}$erenkov counter (FAC) with an aerogel refractive 
index of 1.035. 
The energy threshold of the discriminator for the FAC 
was set higher than the single photon signal in order to suppress 
the trigger rate. 
This was necessary since the $K^{-}$ beam produced a strong 
contamination due to $\delta$ rays in particular at forward angles. 
However, the higher threshold decreased the detection efficiency 
of the FAC to 91 \% for $\pi^{-}$ particles. 
Scattered $\pi^{-}$ particles were bent vertically and analyzed by the 
48D48 spectrometer with five drift chambers \cite{Stotzer1,Khaustov1}. 
The efficiency of track reconstruction was 61 \% for the 
($K^{-}$, $\pi^{-}$) reaction. 
Scintillator hodoscopes defined the acceptance which 
was $-$8$^{\circ}$ to 8$^{\circ}$ in the horizontal direction 
and $-$16$^{\circ}$ to 0$^{\circ}$ in the vertical direction. 
In Fig.\ref{fig:acc}, the acceptance obtained for the 
$^{13}$C($K^{-}$, $\pi^{-}$) reaction is shown. 
There are no unexpected structures due to the inefficiency of 
detectors. 
The large acceptance was essential to excite the doublet states 
simultaneously, which enabled us to derive 
the splitting with a small systematic error. 
Time-of-flight \cite{Sum1} was only used to remove the $K^{-}$ beam 
background. 

The average live time of the DAQ system was 78 \%. 
The $^{13}$C($K^{-}$, $\pi^{-}$) data were accumulated by using 
1.4$\times$10$^{10}$ $K^{-}$ beam particles in total.

\subsection{Energy calibration of the NaI detectors}
\label{sec:level8}

The energy calibration of NaI detectors is particularly important for 
the precise measurement of $\gamma$ rays. 
For the energy calibration in the low energy region, we used 
$\gamma$ rays from $^{22}$Na sources (0.511, 1.275 MeV). 
A $^{22}$Na source was sandwiched between two small plastic 
scintillators to provide a light signal for a trigger by 
its $\beta$-decay, and they were attached to a 3/8$''$ PM tube. 
For the energy calibration in the high energy region near 11 MeV, 
the $^{58}$Ni(n, $\gamma$)$^{59}$Ni reaction was used. 
When a neutron emitted by a $^{244}$Cm-$^{13}$C source thermalizes in 
a moderator and is captured by a $^{58}$Ni nucleus, a 
$\gamma$ ray (8.999 MeV) from the neutron threshold to the 
GS in $^{59}$Ni is emitted. 
A typical energy spectrum of a central NaI detector obtained in 
the energy calibration run using the 
$^{58}$Ni(n, $\gamma$)$^{59}$Ni reaction is shown in 
Fig.\ref{fig:nienergy} where only single hit events, which means that 
there was no signal in neighboring detectors, were selected to make 
full energy peaks dominant. 
A full energy peak at 9.0 MeV is observed separated from the 8.5 MeV 
peak which consists of a single escape peak and $\gamma$ rays from 
the neutron threshold to the state at E$_{x}$=0.465 MeV in $^{59}$Ni. 
Since the expected energy of the reaction $\gamma$ rays is close to 
11 MeV, the systematic error coming from the uncertainty in the 
energy extrapolation from 9.0 MeV to 11 MeV is small. 
The largest peak at 6.1 MeV, which was also used for the 
energy calibration, originates from $\gamma$ rays emitted by the 
$^{13}$C($\alpha$, n)$^{16}$O$^{*}$ reaction in the 
$^{244}$Cm-$^{13}$C source. 
$\gamma$ rays with energies of 8.999, 6.129, 1.275, and 0.511 MeV 
and a pedestal were used for the energy calibration under beam-off 
conditions, and a fit to the five data points with the same weight 
was successfully performed with a linear function for each NaI 
detector. 

Under beam-on conditions, the energy calibration of the whole system was 
monitored by $^{22}$Na sources for stability over more than a few days. 
In addition, LED light fed into all NaI detectors was used to monitor 
the stability over short time durations. 
The gain shift during a beam spill was less than one percent in the 
worst case, which was acceptable for the present measurement. 

For $\gamma$ rays from the $\Lambda$ bound region of 
$^{13}_{\Lambda}$C, a Doppler shift correction was performed. 
A recoil momentum vector of a $^{13}_{\Lambda}$C nucleus was obtained 
from the momentum and direction of incoming $K^{-}$ and 
outgoing $\pi^{-}$ particles. 
The direction of an emitted $\gamma$ ray was calculated from a 
reconstructed vertex of the ($K^{-}$, $\pi^{-}$) reaction and the 
position of the NaI crystal which had the maximum energy deposit. 

As the final energy calibration, we used $\gamma$ rays with energies 
of 4.438 and 15.100 MeV from $^{12}$C nuclei excited by 
the quasi-free ($K^{-}$, $\pi^{-}$) reaction, where the struck neutron 
becomes a $\Lambda$ particle and comes out of the nucleus freely. 
The energy calibration using known $\gamma$ rays 
simultaneously measured by the 
$^{13}$C($K^{-}$, $\pi^{-}$)$^{13}_{\Lambda}$C reaction enabled us 
to determine $\gamma$ ray energies with small systematic errors. 
The excitation energy of a state was obtained from the $\gamma$ ray 
energy by correcting for the recoil of $^{13}_{\Lambda}$C hypernuclei 
due to emitting $\gamma$ rays. 

\section{RESULTS}
\label{sec:level9}
\subsection{Event selection}
\label{sec:level10}

Fig.\ref{fig:ex}(a) shows a 2-dimensional spectrum derived from 
about 10 \% of all data taken at 
0$^{\circ}$$<$$\theta_{\pi}$$<$16$^{\circ}$. 
It shows the momentum of the outgoing particles versus the lab 
scattering angle before event selection. 
The main feature is due to $K^{-}$ beam particles that fired the FAC 
by $\delta$ rays and were not rejected by the hardware trigger. 
Other loci correspond to $K^{-}$ in-flight decays. 
The upper decay locus is 
$K^{-}$$\rightarrow$$\mu^{-}$+$\bar{\nu}_{\mu}$ ($K_{\mu 2}$, 63.5 \%) 
and the lower is 
$K^{-}$$\rightarrow$$\pi^{-}$+$\pi^{0}$ ($K_{\pi 2}$, 21.2 \%). 
Since the ($K^{-}$, $\pi^{-}$) reaction is strongly contaminated 
by the background described above, 
$^{13}$C($K^{-}$, $\pi^{-}$)$^{13}_{\Lambda}$C events are difficult 
to be clearly observed without the event selection. 
The $^{13}$C($K^{-}$, $\pi^{-}$)$^{13}_{\Lambda}$C events are observed 
as a small bump on the huge background in Fig.\ref{fig:ex}(b) which 
is an excitation energy spectrum of $^{13}_{\Lambda}$C. 
The excitation energy was calibrated by using the 
$K^{-}$ decay kinematics. 

In the first stage, events with at least one 1$\sim$20 MeV 
$\gamma$ ray, which amounted to about 7 \% of all data, were selected. 
After the first stage the $K_{\mu 2}$ decay and $K^{-}$ beam events 
were suppressed. 
But the $K_{\pi 2}$ decay still remained, as shown in  
Fig.\ref{fig:ex}(c and d), because a $\pi^{0}$ particle decays 
mainly by emitting two $\gamma$ rays. 

In the second stage, event selection by using the DEC and the 
scintillating target outputs was performed to remove most of the 
remaining background. 
Events with an energy deposit above threshold either in the DEC or in 
the scintillating target were selected. 
The energy threshold of the DEC was set at 2.6 MeV which was lower 
than the minimum ionization, and that of the scintillating target 
was set at 1.2 times higher than the minimum ionization peak. 
This selection tagged 89 \% of $^{13}_{\Lambda}$C production events and 
suppressed the $K^{-}$ decays by 90 \%. 
In addition, outgoing $K^{-}$ particles were removed by using 
time-of-flight. 
Loose vertex cuts of the ($K^{-}$, $\pi^{-}$) reaction 
for the x and y directions were performed, but that for the z direction 
was not done because it might have caused low efficiency 
at forward angles. 
After the second stage a locus corresponding to 
$^{13}$C($K^{-}$, $\pi^{-}$)$^{13}_{\Lambda}$C events 
is clearly observed, and most of the 
$K^{-}$ decays disappeared as shown in Fig.\ref{fig:ex}(e). 
In Fig.\ref{fig:ex}(f) a quasi-free peak is clearly observed, 
although the spectrometer's energy resolution of about 
15 MeV (FWHM) could not resolve excited states. 
The excitation energy of 0 to 25 MeV was selected to purify 
$\gamma$ rays from the $\Lambda$ bound region (0$\sim$11.7 MeV), 
and the energy region of 30 to 100 MeV was regarded as the quasi-free 
region to observe $\gamma$ rays from $^{12}$C nuclei. 

\subsection{Low energy $\gamma$ rays}
\label{sec:level11}

Fig.\ref{fig:lowgamma} shows $\gamma$ ray spectra in the low energy 
region obtained in coincidence with the $^{13}$C($K^{-}$, $\pi^{-}$) 
reaction at 0$^{\circ}$$<$$\theta_{\pi}$$<$16$^{\circ}$, where 
(a) shows $\gamma$ rays from the quasi-free region and 
(b) shows those from the $\Lambda$ bound region. 
The Doppler shift due to the recoil of a $^{13}_{\Lambda}$C nucleus 
was corrected for event-by-event only in Fig.\ref{fig:lowgamma}(b). 

In Fig.\ref{fig:lowgamma}(a), a peak at around 4.5 MeV, which 
corresponds to $\gamma$ rays from the first 2$^{+}$ state in $^{12}$C, 
is the strongest feature. 
The 2$^{+}$ state in $^{12}$C is considered to be copiously produced 
by the $\Lambda$ escape from highly excited states 
in $^{13}_{\Lambda}$C. 
It is well understood that neutron pickup reactions, 
such as the $^{13}$C(p, d) reaction \cite{Hosono1}, 
strongly excite the 2$^{+}$ state. 
As a result of fitting, the peak position was obtained as 
4.467$\pm$0.005(stat) MeV which was shifted to higher energy by 
29 keV. 
A rate-dependent gain shift is believed to be the main reason for 
the apparent energy shift. 
We used the energy shift for the energy calibration to obtain the 
correct energy of $\gamma$ rays from $^{13}_{\Lambda}$C hypernuclei. 
The width of the 2$^{+}$ peak is 240$\pm$10(stat) keV (FWHM) which 
is reasonable. 
Other peak structures are thought to be due to $\gamma$ rays from 
$^{27}$Si or $^{15}$O nuclei produced by the quasi-free process. 

In Fig.\ref{fig:lowgamma}(b), a peak is observed at 4.9 MeV. 
Around this region the 3/2$^{+}$ and 5/2$^{+}$ states, which have 
a configuration of [$^{12}$C(2$^{+}$) $\otimes$ $s_{\Lambda}$], are 
expected to occur. 
Millener predicted that the splitting between the states should be 
74 keV \cite{Millener0}. 
The 3/2$^{+}$ and 5/2$^{+}$ states are excited via $\Delta \ell$=1 
and $\Delta \ell$=3 transitions, respectively, 
by the ($K^{-}$, $\pi^{-}$) reaction. 
Accordingly, the yield of $\gamma$ rays from the 3/2$^{+}$ state to 
the GS must be much greater than that from the 5/2$^{+}$ state at 
0$^{\circ}$$<$$\theta_{\pi}$$<$16$^{\circ}$. 
The 2$^{+}$ peak at around 4.5 MeV is still observed because of 
insufficient selection of the $\Lambda$ bound region. 
A fit to the spectrum was performed using a function consisting 
of two Gaussians and a linear background in the region from 
1.3 to 7.8 MeV. 
As a result of the fitting and the final energy calibration explained 
in the end of Sec.\ref{sec:level8}, the excitation energy of the 
3/2$^{+}$ state in $^{13}_{\Lambda}$C was found to be 
4.880$\pm$0.010(stat)$\pm$0.017(syst) MeV. 
Most of the systematic error originated from the uncertainty of the 
energy calibration using $\gamma$ rays from $^{12}$C nuclei. 
The present measurement is consistent with the excitation energy 
of E$_{x}$=4.89$\pm$0.07 MeV (preliminary) obtained by the 
$^{13}$C($\pi^{+}$, $K^{+}$)$^{13}_{\Lambda}$C 
experiment \cite{Hashimoto1}. 
The width of the 3/2$^{+}$ peak was 220$\pm$25(stat) keV (FWHM) which is 
the same as the width of the 2$^{+}$ peak in Fig.\ref{fig:lowgamma}(a). 
The Doppler shift correction was typically less than 1 \% of the 
$\gamma$ ray energy, which made the width of the 3/2$^{+}$ peak 
narrower by 60 keV (FWHM). 
This result is consistent with identifying that peak as being due to 
$\gamma$ rays from $^{13}_{\Lambda}$C hypernuclei. 

\subsection{High energy $\gamma$ rays}
\label{sec:level12}

$\gamma$ ray spectra in the high energy region obtained in coincidence 
with the $^{13}$C($K^{-}$, $\pi^{-}$) reaction at 
0$^{\circ}$$<$$\theta_{\pi}$$<$16$^{\circ}$ are shown 
in Fig.\ref{fig:highgamma}, 
where (a) shows $\gamma$ rays from the quasi-free region and (b) 
shows those from the $\Lambda$ bound region. 
The Doppler shift due to the recoil of a $^{13}_{\Lambda}$C nucleus 
was corrected for event-by-event only in Fig.\ref{fig:highgamma}(b). 

In Fig.\ref{fig:highgamma}(a), a dominant peak at 15 MeV, which 
corresponds to $\gamma$ rays from the 1$^{+}$ (T=1) state 
to the GS in $^{12}$C, is observed. 
The 1$^{+}$ state in $^{12}$C is frequently produced by $\Lambda$ 
escape from highly excited states in $^{13}_{\Lambda}$C. 
A fit to the spectrum using the energy resolution assumed by 
the GEANT simulator resulted in a peak 
energy of 15.289$\pm$0.022(stat) MeV which was shifted to higher 
energy by 189 keV. 
The energy shift of the peak was also used for the $\gamma$ ray 
energy calibration. 
The ratio between the yields of $\gamma$ rays from the 2$^{+}$ and 
1$^{+}$ states in $^{12}$C is roughly 2:1. 
This value is not inconsistent with the strength ratio of the states 
excited by the $^{13}$C(p, d) reaction \cite{Lewis1}, which 
suggests that a quasi-free knockout of a $\Lambda$ particle is 
dominant in highly excited regions. 

In Fig.\ref{fig:highgamma}(b), a single peak is clearly observed at 
11 MeV, and a small bump, which is due to $\gamma$ rays from 
$^{12}$C nuclei, is also observed at 15 MeV. 
$\gamma$ rays from the 1/2$^{-}$ and 3/2$^{-}$ states in 
$^{13}_{\Lambda}$C are expected to have almost similar yields by 
considering theoretical differential cross sections and the 
acceptance of the spectrometer estimated by the GEANT simulator. 
However, there is no other prominent peak than that at 11 MeV. 
It is natural to think that the observed single peak might include 
$\gamma$ rays from both the 1/2$^{-}$ and 3/2$^{-}$ states, 
although this consideration is inconsistent with a previous result 
that a peak structure, which was considered to be the 3/2$^{-}$ state, 
was observed at E$_{x}$=9.92$\pm$0.13(stat)$\pm$0.5(syst) 
MeV (preliminary) \cite{Hashimoto1}. 
The energy resolution and strength ratio between the full energy and 
single escape peaks are estimated to be 0.35 MeV (FWHM) and 2:1, 
respectively, for 11 MeV $\gamma$ rays. 
The width of the observed single peak seems quite consistent with the 
instrumental peak width, which means that the 1/2$^{-}$ and 3/2$^{-}$ 
states are close to each other. 

\subsection{Splitting of the 1/2$^{-}$ and 3/2$^{-}$ states} 
\label{sec:level13}

As explained in Sec.\ref{sec:level6}, the angular distributions of the 
$^{13}$C($K^{-}$, $\pi^{-}$)$^{13}_{\Lambda}$C reaction selectively 
excite either the 1/2$^{-}$ or 3/2$^{-}$ states even if the states 
are not separated in the energy spectrum of 
$\gamma$ rays. 
It is one of the important advantages of our experiment. 
The events shown in Fig.\ref{fig:highgamma}(b) were 
sub-divided into three 
spectra in Fig.\ref{fig:3angles}, where the scattering angles of 
(a) 0$^{\circ}$$<$$\theta_{\pi}$$<$7$^{\circ}$, 
(b) 7$^{\circ}$$<$$\theta_{\pi}$$<$10$^{\circ}$, and 
(c) 10$^{\circ}$$<$$\theta_{\pi}$$<$16$^{\circ}$ were selected. 
Full energy peaks at 11 MeV are distinctly observed in all spectra. 

A response function of the NaI detectors for 11 MeV $\gamma$ rays was 
obtained by the GEANT simulator which included the detector 
geometry and a procedure for adding the energies of the NaI detectors. 
The response function was folded into the energy resolution of 
0.35 MeV (FWHM) which was estimated by assuming a $\sqrt{E_{\gamma}}$ 
dependence. 
Fits to the spectra were performed in the region from 7.5 to 14 MeV 
using an exponential function for the continuous background and the 
response function for the peak structures. 
As a result of the fittings, peak positions of the $\gamma$ rays were 
obtained as 
11.103$\pm$0.029 MeV, 
11.016$\pm$0.024 MeV, and 
10.980$\pm$0.032 MeV at 
0$^{\circ}$$<$$\theta_{\pi}$$<$7$^{\circ}$, 
7$^{\circ}$$<$$\theta_{\pi}$$<$10$^{\circ}$, and 
10$^{\circ}$$<$$\theta_{\pi}$$<$16$^{\circ}$, respectively. 
The errors are statistical only. 
The fittings gave $\chi^{2}$/N=1.27, 0.88, and 0.87, respectively. 
The surplus at around 10 MeV in Fig.\ref{fig:3angles}(a) made 
the $\chi^{2}$/N worse. 
However, the influence of the surplus on the result of the peak 
position must be little because it is far enough from the peak. 
The excitation energies and systematic errors of the 1/2$^{-}$ and 
3/2$^{-}$ states are derived in Sec.\ref{sec:level14}. 

The splitting of the 1/2$^{-}$ and 3/2$^{-}$ states was 
extracted from Fig.\ref{fig:split}, where the peak positions of 
$\gamma$ rays are plotted as a function of calculated yield ratio, 
\begin{equation}
R=(N(1/2^{-})-N(3/2^{-}))/(N(1/2^{-})+N(3/2^{-})). 
\end{equation}
N(1/2$^{-}$) and N(3/2$^{-}$) stand for the yields of $\gamma$ rays 
from the 1/2$^{-}$ and 3/2$^{-}$ states, respectively. 
The yields were calculated using the theoretical differential cross 
sections of the two states shown 
in Fig.\ref{fig:cross} and the acceptance of the spectrometer. 
The solid angles of the angular regions were 
15.0 msr (0$^{\circ}$$<$$\theta_{\pi}$$<$7$^{\circ}$), 
14.5 msr (7$^{\circ}$$<$$\theta_{\pi}$$<$10$^{\circ}$), 
and 24.1 msr (10$^{\circ}$$<$$\theta_{\pi}$$<$16$^{\circ}$). 
The right, center, and left closed circles indicate the peak 
positions measured at 
0$^{\circ}$$<$$\theta_{\pi}$$<$7$^{\circ}$, 
7$^{\circ}$$<$$\theta_{\pi}$$<$10$^{\circ}$, and 
10$^{\circ}$$<$$\theta_{\pi}$$<$16$^{\circ}$, respectively. 
A fit to the data points was performed with a linear function by 
considering the statistical errors indicated by bars. 
As a result of the fitting, a linear function gave 
$\Delta$E(1/2$^{-}-$3/2$^{-}$) = 
+152$\pm$54(stat) keV for the splitting of 
$^{13}_{\Lambda}$C. 
This splitting will broaden the peak width of the 11 MeV 
$\gamma$ rays by less than 5 \%, which justifies the fitting with 
the response function of a single $\gamma$ ray for each spectrum 
in Fig.\ref{fig:3angles}. 

\section{DISCUSSION}
\label{sec:level14}

We obtained the splitting of the 1/2$^{-}$ and 3/2$^{-}$ states 
as 152$\pm$54(stat) keV. 
Different sources of the systematic error for the splitting are 
discussed below and their contributions are summarized 
in Table \ref{table1}. 

In this analysis, we relied on theoretical differential cross 
sections. 
However the $\gamma$ ray yield at each scattering angle was 
not completely consistent with the calculation. 
The $\gamma$ ray yields in the peak region were 
164$\pm$18 (0$^{\circ}$$<$$\theta_{\pi}$$<$7$^{\circ}$), 
166$\pm$18 (7$^{\circ}$$<$$\theta_{\pi}$$<$10$^{\circ}$), and 
142$\pm$21 (10$^{\circ}$$<$$\theta_{\pi}$$<$16$^{\circ}$). 
Whereas the theoretically expected yields were 
385 (0$^{\circ}$$<$$\theta_{\pi}$$<$7$^{\circ}$), 
167 (7$^{\circ}$$<$$\theta_{\pi}$$<$10$^{\circ}$), and 
179 (10$^{\circ}$$<$$\theta_{\pi}$$<$16$^{\circ}$). 
If we assume that these inconsistencies originate from uncertainties 
in the theoretical differential cross sections, R in 
Fig.\ref{fig:split} would vary as 
0.77$^{+0.23}_{-0.31}$ (0$^{\circ}$$<$$\theta_{\pi}$$<$7$^{\circ}$), 
0.01$^{+0.01}_{-0.01}$ (7$^{\circ}$$<$$\theta_{\pi}$$<$10$^{\circ}$), 
and 
$-$0.82$^{+0.05}_{-0.18}$ (10$^{\circ}$$<$$\theta_{\pi}$$<$16$^{\circ}$). 
The uncertainty of R produced a systematic error of 30 keV as the 
maximum deviation from the central value for the splitting. 
This systematic error is the largest one. 
All possible causes of the inconsistent $\gamma$ ray yield were 
investigated. 
But an inconsistency of about a factor of 2 remained. 
The theoretical differential cross section of the 1/2$^{-}$ state 
at the forward angles is especially sensitive to the $K^{-}$ beam 
momentum of around 0.9 GeV/c. 
For example, the differential cross section is expected to decrease by 
a factor of 3 at 1.0 GeV/c \cite{Motoba1}. 
There is a possibility that a small ambiguity of the $K^{-}$ beam 
momentum in previous ($K^{-}$, $\pi^{-}$) experiments, on which 
differential cross sections were adjusted theoretically, might 
produce such a large inconsistency at the forward angles. 
If it is a cause of the inconsistency, the systematic error would be 
much smaller. 

Several fittings with different functions, such as 2 Gaussians and 
a linear background, and widths were applied to study how the 
choice of functions can influence the splitting, and 
a systematic error originating from choosing different fitting 
functions was estimated to be 19 keV at most. 

The influence of the Doppler shift correction was also studied by 
using the GEANT simulator, and it was found to be negligible because 
the NaI detectors installed symmetrically in the vertical direction 
almost canceled it. 
The energy calibration of the NaI detectors affects 
the splitting very little because $\gamma$ rays from both 
the states were measured simultaneously. 

The total systematic error of the splitting was estimated to be 
36 keV. 
The final result of the splitting was determined to be 
$\Delta$E(1/2$^{-}-$3/2$^{-}$) = 
+152$\pm$54(stat)$\pm$36(syst) keV. 
The $\ell s$-splitting of a nucleon in the $p$-orbit around this 
mass region is 3$\sim$5 MeV, thus the $\ell s$-splitting of a $\Lambda$ 
particle in the $p$-orbit is about 20$\sim$30 times smaller. 
After the final energy calibration explained in the end of 
Sec.\ref{sec:level8}, the excitation energies of the 1/2$^{-}$ and 
3/2$^{-}$ states were obtained as 
10.982$\pm$0.031(stat)$\pm$0.056(syst) MeV and 
10.830$\pm$0.031(stat)$\pm$0.056(syst) MeV, 
respectively. 
The uncertainty of the energy calibration using $\gamma$ rays from 
$^{12}$C nuclei and the choice of fitting functions were mainly 
considered to estimate the systematic errors. 
The $j_{\Lambda}$=$\ell_{\Lambda}-$1/2 (($p_{1/2}$)$_{\Lambda}$) state 
appears higher in energy, as in normal nuclei, which is consistent with 
theoretical predictions \cite{Itonaga1,Hiyama1}. 
The present measurement is consistent with a $\gamma$ ray energy 
of 10.95$\pm$0.1(stat)$\pm$0.2(syst) MeV in Ref.\cite{May3}. 

Recently $YN$ interactions were refined in both the OBE model 
\cite{Rijken1} and the quark model \cite{Fujiwara1,Kohno1}. 
The strength of a $\Lambda$-nucleus $ALS$ force is different 
between the models. 
According to calculations by Hiyama $et$ $al$. in the framework of 
the microscopic 3$\alpha$+$\Lambda$ model, the $ALS$ force 
of the OBE models decreased the splitting of $^{13}_{\Lambda}$C by 
only 20$\sim$30 \% \cite{Hiyama1}. 
The predicted splittings were 0.75, 0.96, and 0.39$\sim$0.78 MeV 
by means of Nijmegen model D, Nijmegen model F, and Nijmegen 
soft-core models (a$\sim$f), respectively. 
These calculations systematically show that the $YN$ interaction given 
by the OBE models predicts larger $\ell s$-splittings than the present 
measurement. 
The state-of-the-art calculation of the $YN$ interaction 
based on the OBE models is unable to reproduce the present result. 
On the other hand, in a calculation of the strength of the one-body 
$\ell s$-force starting from a quark-based $YN$ interaction, 
the strength of the $ALS$ force amounted to approximately 85 \% of 
that of the $\ell s$-force \cite{Kohno1}. 
A calculation using a large $ALS$ force based on the quark model 
gave about 0.2 MeV for the splitting of 
$^{13}_{\Lambda}$C \cite{Hiyama1}. 
The difference between these results was mainly due to different 
strengths of the $ALS$ force. 

Since the splitting for $^{13}_{\Lambda}$C is very small, forces 
besides the $\ell s$-force may also contribute to the splitting. 
It has been pointed out by Millener that a tensor 
force makes a significant contribution to the splitting of 
$^{13}_{\Lambda}$C \cite{Millener1}. 
The nuclear $\ell s$-force mixes a small S=1 component into 
the $^{12}$C core wave function, and other forces arise from 
the S=1 component. 
His prediction for the splitting of $^{13}_{\Lambda}$C was 107 keV, 
where the spin-spin force (+42 keV), the $\ell s$-force 
(+280 keV), and the tensor force ($-$215 keV) were considered. 
The result almost reproduces the present measurement. 

A systematic study of light $\Lambda$-hypernuclei shows that the 
$YN$ interactions based on the OBE model need to be modified so that 
a smaller $\ell s$-splitting, which has been indicated by the present 
experiment, can be accommodated \cite{Millener0}. 
A new mechanism will be required for the unified understanding of 
the baryon-baryon ($NN$, $YN$ and $YY$) interaction. 

\section{SUMMARY}
\label{sec:level15}

We performed the 
$^{13}$C($K^{-}$, $\pi^{-}\gamma$)$^{13}_{\Lambda}$C experiment at 
0.93 GeV/c at the AGS of BNL to obtain the 
$\ell s$-splitting of $\Lambda$ single particle states in 
$^{13}_{\Lambda}$C with high precision. 
We succeeded in measuring $\gamma$ rays from the 1/2$^{-}$ and 
3/2$^{-}$ states, which have predominantly a 
[$^{12}$C$_{g.s.}$(0$^{+}$) $\otimes$ $p_{\Lambda}$] 
configuration, to the GS in $^{13}_{\Lambda}$C by using NaI detectors. 
The splitting was found to be 
$\Delta$E(1/2$^{-}-$3/2$^{-}$) = +152$\pm$54(stat)$\pm$36(syst) keV 
which was almost 20$\sim$30 times smaller than that of single particle 
states in nuclei around this mass region. 
The excitation energies of the 1/2$^{-}$ and 3/2$^{-}$ states were 
obtained as 10.982$\pm$0.031(stat)$\pm$0.056(syst) MeV and 
10.830$\pm$0.031(stat)$\pm$0.056(syst) MeV, 
respectively. 
The $j_{\Lambda}$=$\ell_{\Lambda}-$1/2 (($p_{1/2}$)$_{\Lambda}$) state 
appeared higher in energy, as in normal nuclei, which is consistent with 
theoretical predictions. 
We also observed $\gamma$ rays from the 3/2$^{+}$ state to 
the GS in $^{13}_{\Lambda}$C, and the excitation energy of 
the state was obtained as 
4.880$\pm$0.010(stat)$\pm$0.017(syst) MeV. 

\acknowledgments

We thank the staff of the Brookhaven AGS for their support in running the 
experiment. 
We thank Prof. T. Motoba and Prof. K. Itonaga for calculations of 
angular distributions. 
We thank Ms. Elinor Norton in the Chemistry Department at BNL for 
determining the $^{13}$C enrichment in our benzene target. 
We thank Prof. D. Alburger for the careful reading and correcting 
of the manuscript. 
We are grateful to Prof. K. Imai for his continuous support to the 
experiment. 
This experiment is financially supported in part by the Grant-in-Aid 
for Scientific Research in Priority Areas (Strangeness Nuclear Physics) 
08239205 and for International Scientific Research 10044086 and in 
part by the U.S. Department of Energy under contract 
No. DE-AC02-98CH10886. 


%
%

\begin{figure}
\begin{center}
\caption{Theoretically calculated differential cross sections of the 
1/2$^{-}$ (solid curve) and 3/2$^{-}$ (dotted curve) states excited by 
the $^{13}$C($K^{-}$, $\pi^{-}$)$^{13}_{\Lambda}$C reaction at 
0.9 GeV/c. }
\label{fig:cross}
\end{center}
\end{figure}

\begin{figure}
\begin{center}
\caption{The detector system at the target region is shown 
schematically. See text for a description of each detector element. }
\label{fig:aroundtgt}
\end{center}
\end{figure}

\begin{figure}
\begin{center}
\caption{Acceptance of the 48D48 spectrometer obtained for the 
$^{13}$C($K^{-}$, $\pi^{-}$) reaction. }
\label{fig:acc}
\end{center}
\end{figure}

\begin{figure}
\begin{center}
\caption{A $\gamma$ ray energy spectrum of a central NaI detector 
obtained in an energy calibration run using the 
$^{58}$Ni(n, $\gamma$)$^{59}$Ni reaction. 
A $^{244}$Cm-$^{13}$C source was used as a neutron source. }
\label{fig:nienergy}
\end{center}
\end{figure}

\begin{figure}
\begin{center}
\caption{Momentum vs lab scattering angle spectra (a, c, and e) 
and excitation energy spectra of $^{13}_{\Lambda}$C at 
0$^{\circ}$$<$$\theta_{\pi}$$<$16$^{\circ}$ (b, d, and f) are shown. 
Spectra (a-b), (c-d), and (e-f) are before the 
first stage of event selection (about 10 \% of all data), after 
the first stage, and after the second stage, 
respectively. }
\label{fig:ex}
\end{center}
\end{figure}

\begin{figure}
\begin{center}
\caption{Energy spectra of low energy $\gamma$ rays obtained in 
coincidence with the $^{13}$C($K^{-}$, $\pi^{-}$) reaction at 
0$^{\circ}$$<$$\theta_{\pi}$$<$16$^{\circ}$. 
The quasi-free region was selected in (a), and the $\Lambda$ bound region 
was selected in (b). }
\label{fig:lowgamma}
\end{center}
\end{figure}

\begin{figure}
\begin{center}
\caption{Energy spectra of high energy $\gamma$ rays obtained in 
coincidence with the $^{13}$C($K^{-}$, $\pi^{-}$) reaction at 
0$^{\circ}$$<$$\theta_{\pi}$$<$16$^{\circ}$. 
The quasi-free region was selected in (a), and the $\Lambda$ bound region 
was selected in (b). }
\label{fig:highgamma}
\end{center}
\end{figure}

\begin{figure}
\begin{center}
\caption{Energy spectra of $\gamma$ rays from the $\Lambda$ bound 
region obtained in coincidence with 
the $^{13}$C($K^{-}$, $\pi^{-}$) reaction at 
0$^{\circ}$$<$$\theta_{\pi}$$<$7$^{\circ}$ (a), 
7$^{\circ}$$<$$\theta_{\pi}$$<$10$^{\circ}$ (b), 
and 10$^{\circ}$$<$$\theta_{\pi}$$<$16$^{\circ}$ (c). 
Dashed and dotted lines show response functions and exponential 
backgrounds, respectively. 
Solid lines show the total of them. }
\label{fig:3angles}
\end{center}
\end{figure}

\begin{figure}
\begin{center}
\caption{Peak positions obtained by fitting the $\gamma$ ray spectra 
are shown as a function of yield ratio (R). See text for the 
definition of R. }
\label{fig:split}
\end{center}
\end{figure}

%
%

\begin{table}
\caption{Summary of systematic errors for the splitting. }
\label{table1}
\begin{center}
\begin{tabular}[h]{l c}
Effect & Systematic Error (keV)\\
\tableline
Cross section uncertainty & 30\\
Fits to spectra & 19\\
Doppler shift correction & 1\\
Energy calibration of NaI detectors & 1\\
\tableline
{\bf Total} & {\bf 36}\\
\end{tabular}
\end{center}
\end{table}

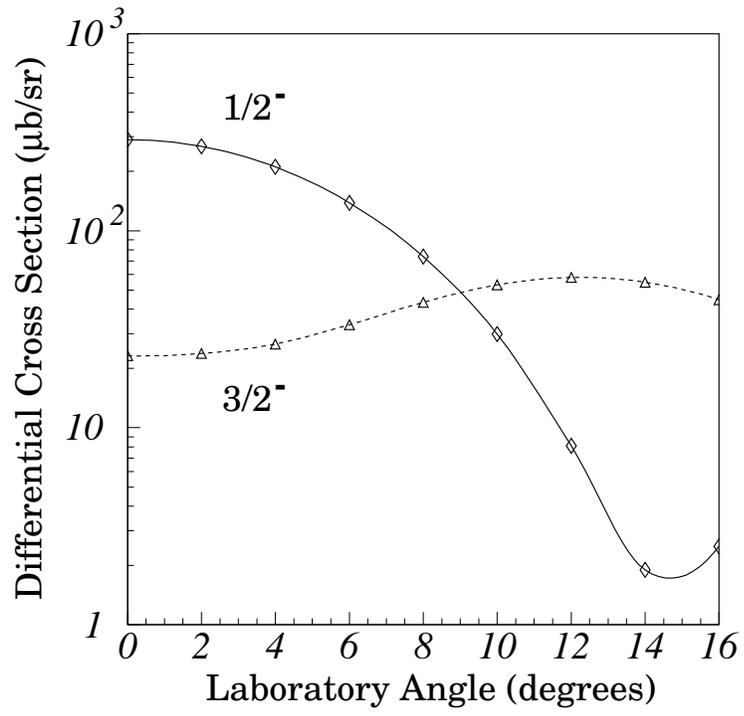
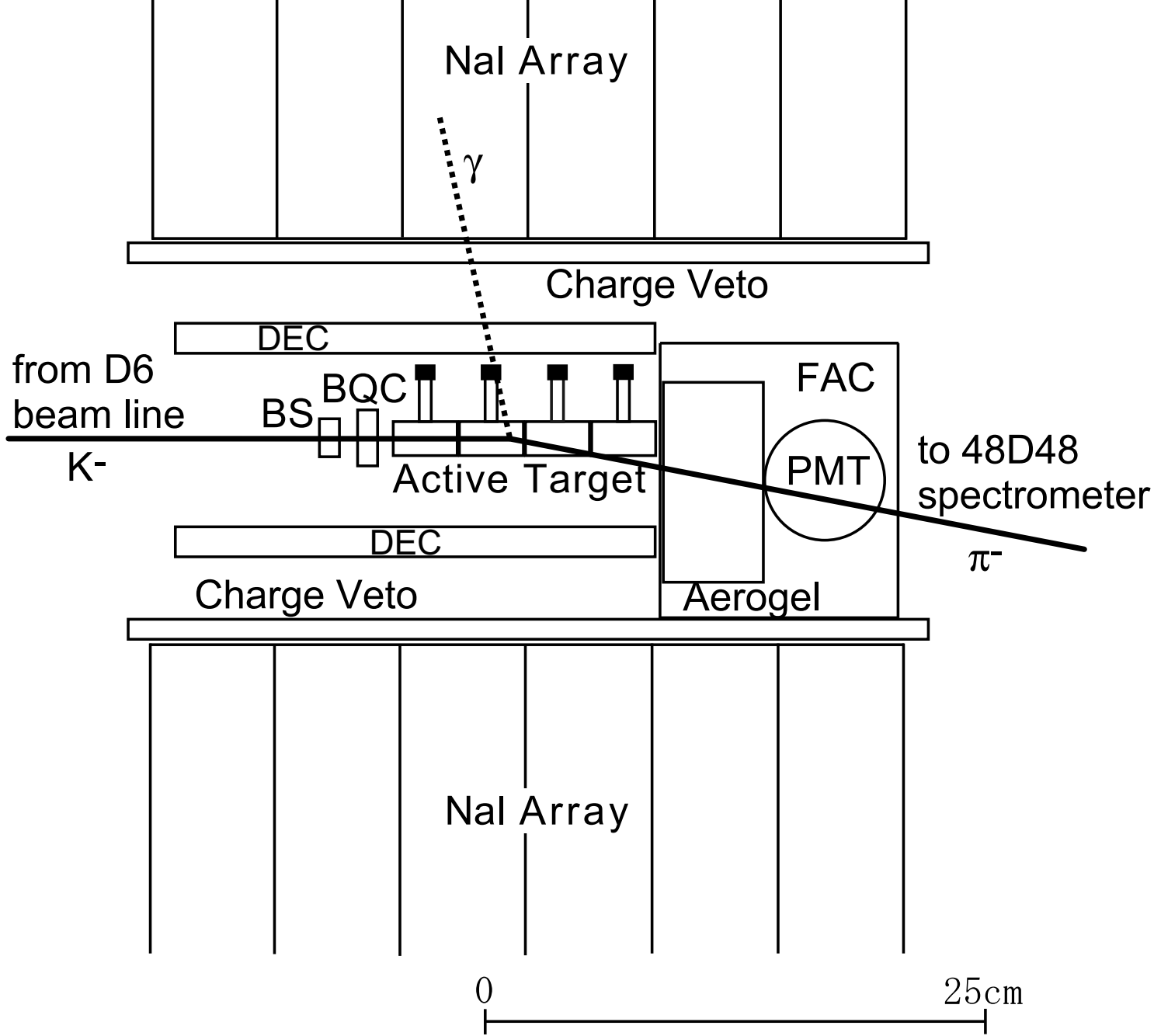
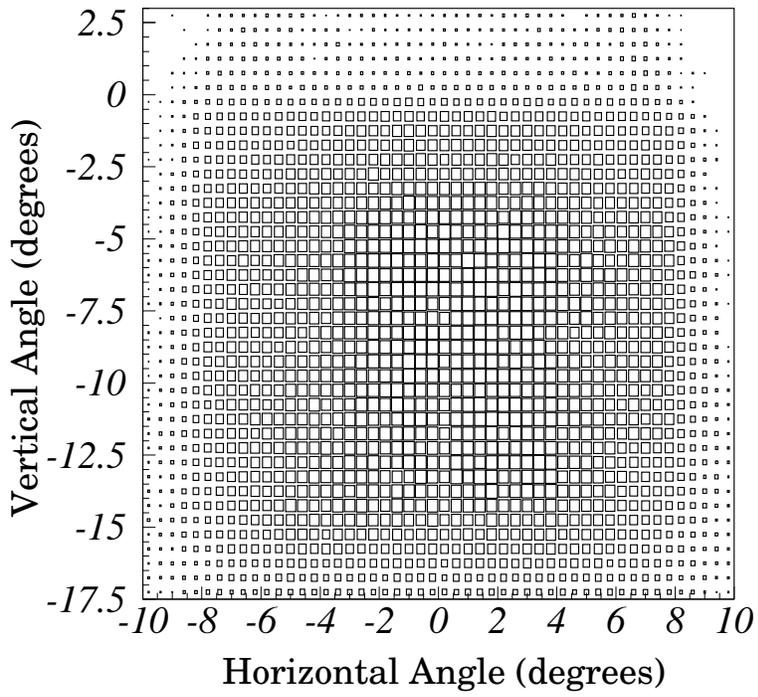
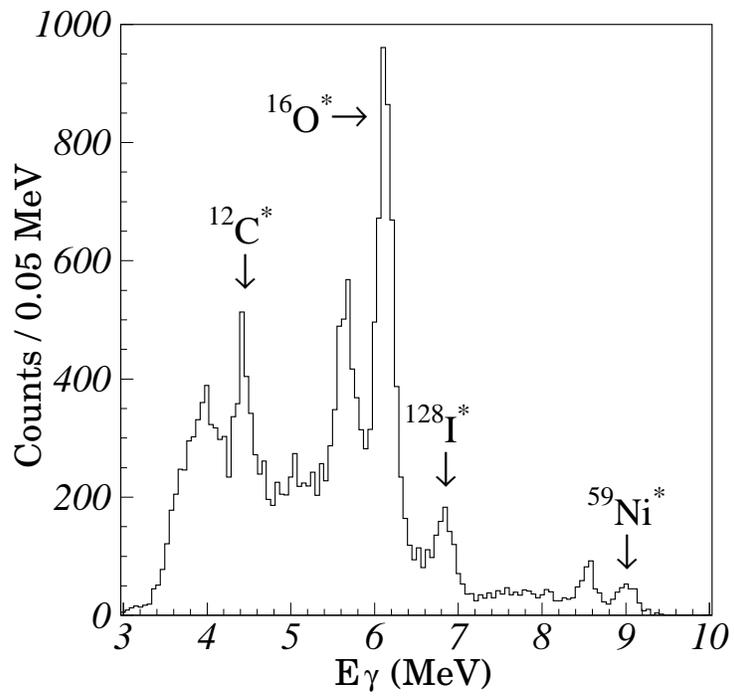
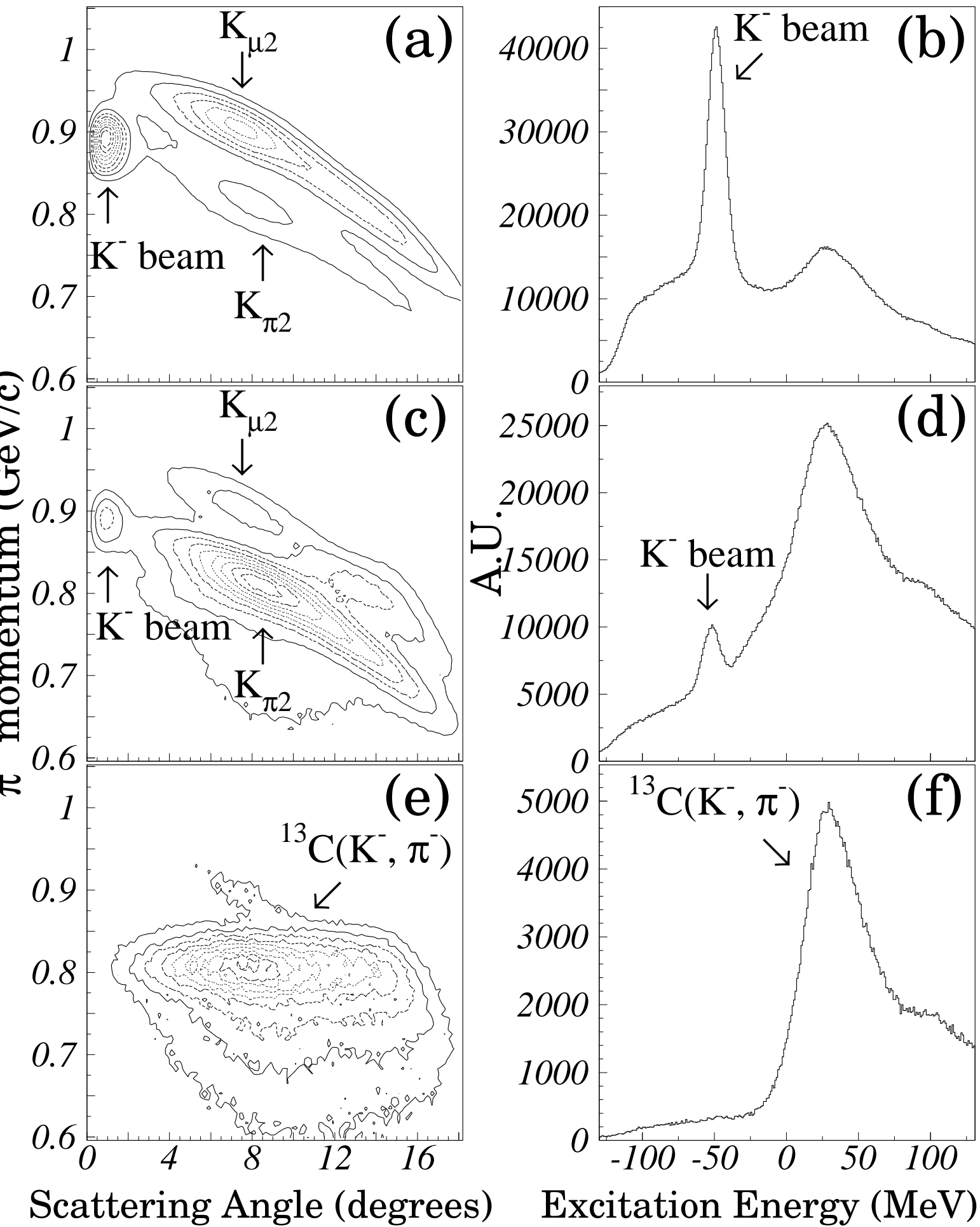
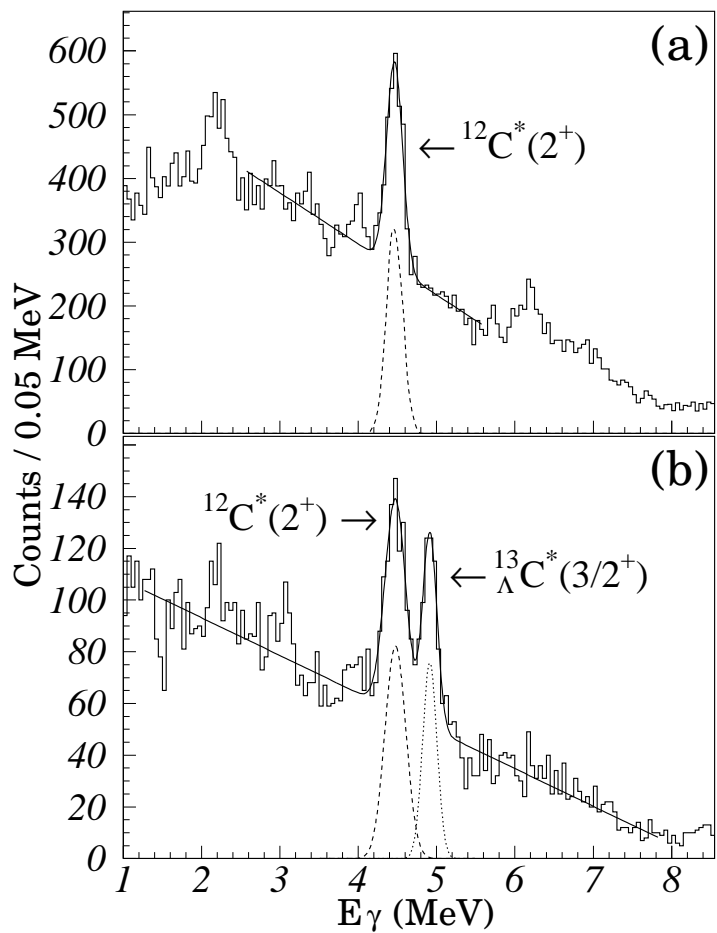
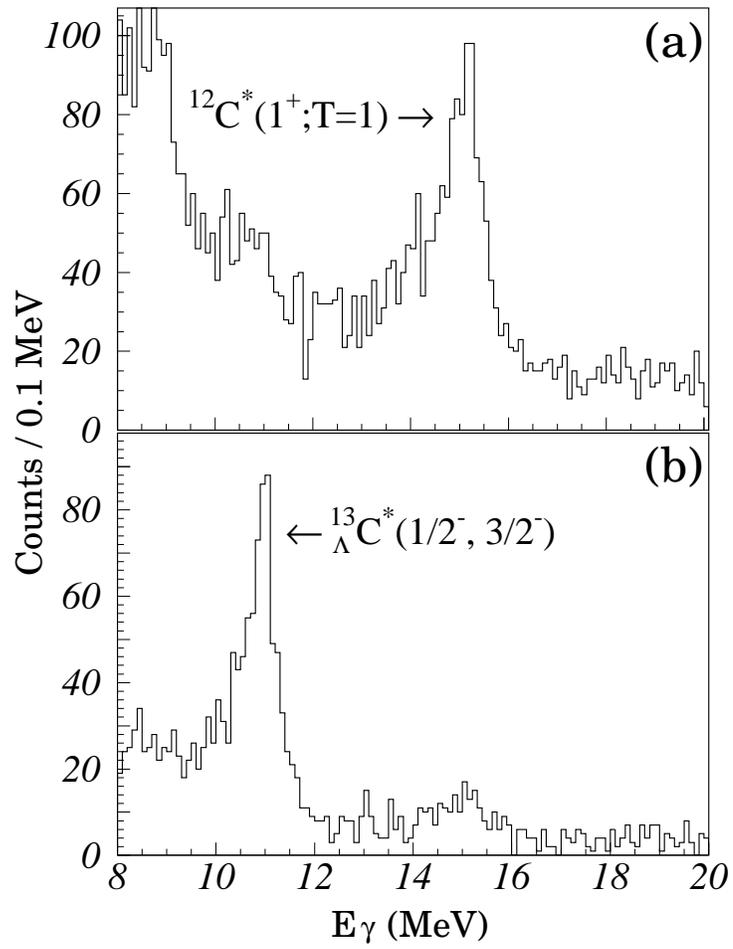
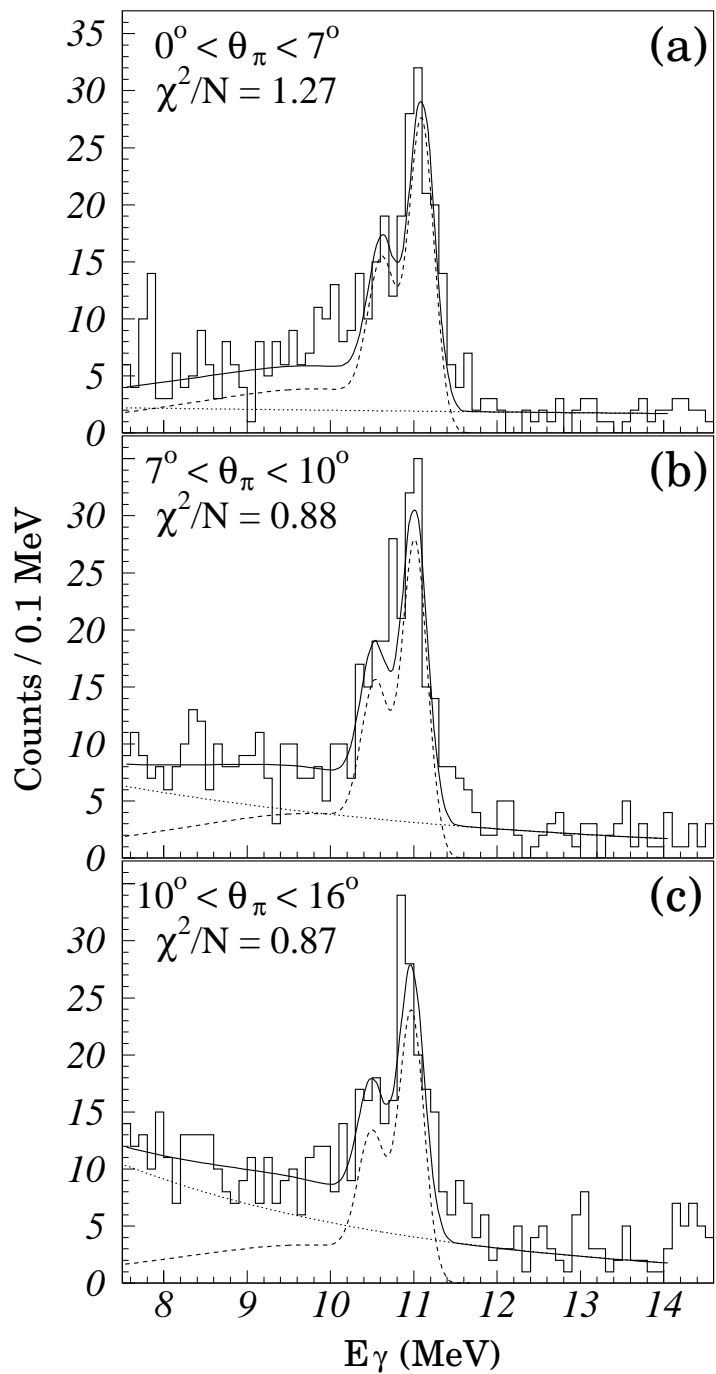
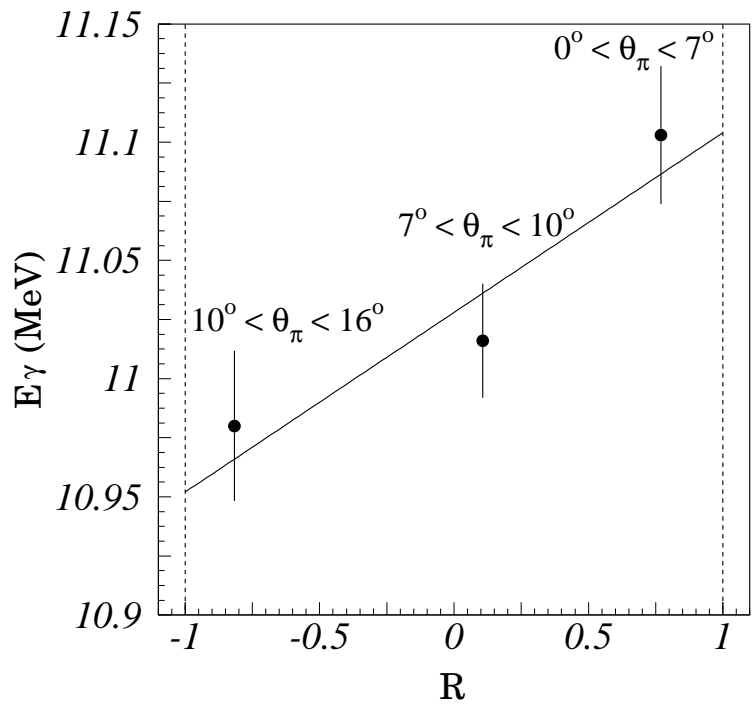
\end{document}